\documentclass[12pt]{article}
\usepackage{amssymb}
\usepackage{amsmath}
\usepackage{graphicx}
\begin{document}

\vspace*{1cm}

\begin{center} 
\setlength{\baselineskip}{24pt}
{\LARGE The Latest on the Sokal Affair: \\
Beyond Three Extremisms
}
\end{center}

\begin{center}
\vspace{2cm}
{\large Basarab Nicolescu}

\vspace{0.5cm} 
 Theory Group,  Laboratoire de Physique Nucl\'eaire  et des Hautes  \'Energies 
(LPNHE)\footnote{Unit\'e  de Recherche des Universit\'es 
  Paris 6 et Paris 7, Associ\'ee ou CNRS},
 CNRS and Universit\'e Pierre et Marie Curie, Paris\\ 
     {\small e-mail: \texttt{nicolesc@lpnhep.in2p3.fr }}

\vspace{3cm}
\textbf{Abstract}
\end{center}
After a short summary of the Sokal Affair and of the point of view of Steven Weinberg, we discuss certain aspects of the scientist ideology present in the last book of Alan Sokal \textsl{Pseudoscience and Postmodernism: Antagonists or Fellow-Travelers?}, published in France. The danger of the three extremisms present in our time - relativist extremism, scientist extremism and its mirror-image, the religious extremism - is underlined. We point out also the necessity of a transdisciplinary dialogue between the different disciplines as a rampart against the fascination caused by these three extremisms.

\newpage

The Sokal affair started with a hoax. In 1994, a mathematical physicist from the University of New York, unknown outside a closed circle of physicists, sent an article to the journal \textit{Social Text}, entitled \textit{Transgressing the Boundaries: Toward a Transformative Hermeneutics of Quantum Gravity} \cite{sokaT}.

The text was peppered with accurate quotations by physicists such as Bohr
and Heisenberg and philosophers, sociologists, historians of science or psychoanalysts such as Kuhn, Feyerabend, Latour, Lacan, Deleuze, Guattari, Derrida, Lyotard, Serres or Virilio. In the bibliography the text also listed authors such as Lupasco: the included middle being given as an example of "feminist logic". Sokal's commentary made however a number of absurd assertions, creating the impression that he was perfectly attuned to postmodernism and, in particular, to the \textit{Cultural Studies} current of thinking. The journal's editors were delighted with the apparent adhesion of a physicist to their cause and published Sokal's text without hesitation, without even the slightest verification. 

Almost as soon as the article appeared in print Sokal himself revealed the hoax in a further article published in \textit{Lingua Franca} \cite{sokaP} and, from that moment on, buoyed by the effect the Internet, his fame was made. On a political level, Sokal wanted to show his friends of the American left wing that a revolution or social transformation could not be carried out based of the notion of reality as presented by philosophical relativism. Only physics (in fact, according to Sokal's view on what physics is) could play such a role.

This unleashed a storm of reactions on the Internet and the publication of books and countless journal articles on the exposure of a very real problem. For some Sokal was the apostle of Enlightenment against the post-modern obscurantists. For others he was part of the thought police or merely a philistine imposture.

The Sokal Affair had the merit of highlighting a phenomenon that is becoming increasingly pervasive in contemporary culture: that of the radicalisation of relativism. Indeed, it is evident that relativist extremism has a nefarious capacity to appropriate the language of the exact sciences. Taken out of context, this language can then be manipulated to say almost anything, indeed to demonstrate that everything is relative. The first victims of this form of deconstruction are the exact sciences that find themselves relegated to the rank of mere social constructs among many others, once the constraint of experimental verification is put into parentheses. It is, therefore, not surprising that in a few months Alan Sokal became the hero of a community that is aware of a blatant contradiction between its every day practices and its social and cultural representation.

Paradoxically, however, the Sokal Affair served to reveal the extent of another form of extremism namely \textit{scientist extremism}, the mirror-image of \textit{religious extremism}. Indeed, Sokal's position was supported by some notable heavyweights including the Nobel Prizewinner Steven Weinberg, who wrote a long article in the \textit{New York Review of Books} \cite{wein96}.

For Weinberg ''The gulf of misunderstanding between scientists and other intellectuals seems to be at least as wide as when C. P. Snow worried about it three decades ago''. But what is the cause of this ``gulf of misunderstanding''? According to Weinberg, one of the essential conditions for the birth of modern science what the severance of the world of physics from the world of culture. Consequently, from that moment on, any interaction between science and culture can only be seen as detrimental. And with this, in one fell swoop, he dismissed as irrelevant the philosophical considerations of the founding fathers of quantum mechanics.

To some, Weinberg's arguments caused some to smart as if from the whiff of the scientism of another century: the appeal to common sense in support the claim about the reality of physical laws, the discovery through physics of the world Ôas it isÕ, the one-to-one correspondence between the laws of physics and Ôobjective realityÕ, the hegemony on an intellectual level of the natural science (because Ôwe have a clear idea of what it means for a theory to be true or falseÉÕ). However, Weinberg is certainly neither a positivist nor a mechanist. He is without doubt one of the most distinguished physicists of the 20th century and a man of broad culture, and due attention should be paid to his arguments.

Weinberg's central idea, repeatedly pounded out like a mantra, concerns the discovery by physics of the existence of \textit{impersonal laws}, the impersonal and eternal laws which guarantee the objective progress of science and which explain the unfathomable gulf between science and culture. The tone of the argument is overtly prophetic, as if in the name of a strange religion without God. One is almost tempted to believe in the immaculate conception of science. We are led to understand, therefore, that for Weinberg, what is really at stake in the Sokal Affair is the \textit{status of truth and reality}. Truth, by definition, cannot depend on the social environment of the scientist. Science is the custodian of truth and, as such, its severance with culture is complete and definitive. There is only one Reality: the objective reality of physics. Weinberg states unequivocally that for culture or philosophy, the difference between quantum mechanics and classical mechanics or between Newton and EinsteinÕs theory of gravitation is insignificant. The contempt with which Weinberg treats the notion of hermeneutics therefore seems quite natural.

Weinberg's conclusion comes down like a cleaver: ''the conclusions of physics may become relevant to philosophy and culture when we learn the origin of the universe or the final laws of nature'', that is to say never!

In 1997, Sokal decided to write a counterargument to the famous \textit{Social Text} article, to express his actual thinking. For this he took on as a co-author, the Belgian physicist Jean Bricmont, who was supposed to have a good knowledge of the intellectual situation in France. The collaboration resulted in the publication of \textit{Intellectual Impostures}, which via its provoking title was intended to become a bestseller. I do not know if its hoped-for destiny was ever attained, but the content of the book surprised by its intellectual patchiness and the repercussions were far beneath the resounding success of Sokal's original hoax. The impostors in question - most of whom are French - are Jacques Lacan, Julia Kristeva, Luce Irigaray, Bruno Latour, Jean Baudrillard, Gilles Deleuze, Felix Guattari and Paul Virilio. 

It would be tedious to go over again, here, the nature of the
``imposture''. The method of the American and Belgian physicists was simple: sentences are removed from their original context and then shown to be meaningless or inaccurate from the perspective of mathematics or physics. For example, Lacan wrote that: ``In this pleasure space (\textit{espace de jouissance}), to take something that is bounded, closed, constitutes a locus, to speak of it is a topology'' \cite{sokaI}. Sokal and BricmontÕs commentary follows: ``In this sentence, Lacan uses four mathematical technical terms (`space', `bounded', `closed' and `topology') but without paying attention to their \textit{meaning}; this sentence is meaningless from a mathematical point of view.'' \cite{idem27}  The method used by the authors disqualifies the book and people thought that the Sokal Affair was definitively closed.

But Sokal is back. He has just published in France a new book with the catchy title \textit{Pseudoscience and Postmodernism: Antagonists or Fellow-Travelers?} \cite{sokaPp}. It is to be noted in passing that Jean Bricmont has written or largely contributed to a third of the book. Indeed, in the introduction, he sets the scene by quoting, in the first lines, Jerry Fodor: ``The point of view to which I adhere [...] is \textit{scientism}.'' 

It is undoubtedly the case that drawing comparisons between postmodernism and pseudoscience has a certain appeal, even if Sokal concedes that he has only found a few postmodern thinkers who rally to the cause of the pseudosciences. 

The real novelty of the book is elsewhere, however, and can be found in
Appendix A entitled `Religion as pseudoscience''. It should be made clear,
here, that Sokal is not referring to cults or new religious movements, but
to established religions : Christianity, Judaism, Islam and Hindouism. It is, therefore, hardly surprising that Sokal points an accusatory finger at Pope John Paul II, described by Sokal as the head of a ``major pseudoscientific cult, Catholicism'' \cite{idem51}.

This last affirmation smacks simply of defamation though it is interesting, nonetheless, to try to understand why religion for Sokal (and Bricmont) is a pseudoscience in the same way that astrology is. 

Sokal candidly explains that ``religion refers to real or alleged phenomena, or real or alleged causal relations, which modern science views as improbable''. He goes on ``that [religion] attempts to base its affirmations on a form of reasoning and system of proofs that is far from satisfying the criteria of modern science in terms of logic and verification'' [8]. Sokal's epistemological error is plain to see: he assumes modern science to be the only arbiter of truth and Reality. At no point does he entertain the idea of a plurality of levels of Reality, with science applying to certain levels and religion to others. For Sokal there can only be one level of Reality, a hypothesis which is epistemologically untenable in light of what modern science has taught us. 

Sokal's attitude is strangely reminiscent of LeninÕs position when, in 1908, in \textit{Materialism and Empiriocriticism}, he attacked the theories of physics implied the multi-dimensional space-time, by proclaiming that the revolution could only be carried out in four dimensions. Lenin, like Sokal, believed in the existence of a single level of Reality. Or rather, he stuck to this belief, to justify his revolution. 

It seems clear that we could find numerous arguments to pick serious holes in the affirmations made by Sokal and his friends. But, to my mind, this approach is futile, and going down this alley merely runs with the risk entering an interminable polemic involving the parodying of the other sideÕs position to the point of trading insults. The three extremisms at stake here, of which religious extremism is the one which the public is most familiar, have brought to the fore a fundamental problem Ð that of the status of truth and Reality Ð and of showing us the consequences, including the political consequences of this problem.

What we need to do at present it to go beyond these three extremisms, which are the germ of new forms of totalitarianism. The Sokal affair gives us a good opportunity to reformulate, on a new and rigorous basis, not only the conditions for the dialogue between the hard sciences and the humanities, but also of the dialogue between science and culture, science and society and science and spirituality. Ultimately, the source of the violent polemic unleashed by the Sokal Affair is the considerable confusion between the \textit{tools} and the \textit{conditions} of the dialogue. Sokal and his friends are right to denounce the anarchic migration of concepts from the hard sciences to the humanities which can only lead to a semblance of rigor and validity. But the moderate relativists are also right to denounce the desire expressed by certain scientists to ban the dialogue between science and culture. Indeed, Weinberg falls prey to exactly the same form of confusion as his contradictors. Why does he declare himself to be ``against philosophy'' (title of one of the chapters of his book \textit{Dreams of a Final Theory} 
\cite{wein92})? Simply because he criticises the tools of philosophy as not being productive in the process of scientific creation. 

Are we obliged to choose between these three extremisms, as the only possible outcome? Certainly not. Transdisciplinarity is a rampart against the fascination caused by the three extremisms. 

If there is a dialogue between the different disciplines, it cannot be founded on the concepts of one or other discipline but what the disciplines have in common: the Subject itself. The Subject which, in the interaction with the Object, refuses any form of formalisation and which always maintains an element of irreducible mystery. The Subject which throughout the 20th century has been considered as the Object: object of experience, object of ideologies which claimed to be scientific, object of ÔscientificÕ studies that are aimed at dissecting it, formalising and manipulating it, and revealing in so doing, a self-destructing process in the irrational struggle of the human being against himself. At the end of the day, it is to the resurrection of the Subject that the Sokal Affair sends us Ð a veritable transdisciplinary quest, the painstaking working out of a new art of living and thinking.

 \end{document}